\documentclass[%
 aip,
 jmp,%
 amsmath,amssymb,
 reprint,%
]{revtex4-1}

\usepackage{graphicx}

\usepackage{dcolumn}
\usepackage{bm}

\usepackage{xspace}

\usepackage{graphicx}
\usepackage{dcolumn}
\usepackage{bm}
\usepackage{color}

\begin{document}

\title{A compact micro-wave synthesizer for transportable cold-atom interferometers}

\author{J. Lautier, M. Lours, and A. Landragin}

\email{arnaud.landragin@obspm.fr}

\affiliation{LNE-SYRTE, Observatoire de Paris, CNRS, UPMC, 61
avenue de l'Observatoire, 75014 Paris, France}


\begin{abstract}
We present the realization of a compact micro-wave frequency synthesizer for an atom interferometer based on stimulated Raman transitions, applied to transportable inertial sensing. Our set-up is intended to address the hyperfine transitions of $^{87}$Rb at 6.8 GHz. The prototype is evaluated both in the time and the frequency domain by comparison with state-of-the-art frequency references developed at LNE-SYRTE. In free-running mode, it features  a residual phase noise level of -65 dBrad$^2$.Hz$^{-1}$ at 10-Hz offset frequency and a white phase noise level in the order of -120 dBrad$^2$.Hz$^{-1}$ for Fourier frequencies above 10 kHz. The phase noise effect on the sensitivity of the atomic interferometer is evaluated for diverse values of cycling time, interrogation time and Raman pulse duration. To our knowledge, the resulting contribution is well below the sensitivity of any demonstrated cold atom inertial sensors based on stimulated Raman transitions. The drastic improvement in terms of size, simplicity and power consumption paves the way towards field and mobile operations.
\end{abstract}

\pacs{Valid PACS appear here}

\maketitle

\section{\label{sec:level1}INTRODUCTION}

Over the past two decades atom interferometry has kept supporting new concepts and experimental set-ups to perform high precision measurements~\cite{Berman}. In particular light pulse atom interferometry using stimulated Raman transitions has been turned into a very powerful tool for both fundamental and applied sciences, especially to measure inertial forces such as the acceleration due to gravity~\cite{Peters}. Since the first realization of such inertial sensors~\cite{Kasevich}, their accuracy and their stability have been continuously improved~\cite{Louchet,Hu}. Instruments of this kind now directly compete with ``classical'' devices~\cite{Merlet}. According to their principle of operation, the sensitivity and the accuracy of the measurement rely on the control of the absolute frequencies and the phase difference of the Raman lasers. Indeed, it has been shown~\cite{Borde} that the atomic phase difference at the interferometer output only depends on the optical phase difference between the two Raman lasers, at the position of the center-of-mass of the matter-wave packets, at the time of the light pulses. In particular, the measurement sensitivity relies on the performance of the microwave reference signal used to control the phase difference between the Raman laser.\\
As such experiments have been transformed into reliable instruments, their field deployment becomes more than relevant~\cite{Barrett}. For example, accelerometers relying on atom interferometry have already demonstrated to be transportable~\cite{BIPM} and robust enough for airborne operations~\cite{Geiger}. However some utilizations for fundamental science~\cite{STE-QUEST} or more applied studies~\cite{nav} would require a higher level of integration. Progress has been made to simplify the sensor head~\cite{Bodard} and the laser system~\cite{Stern}, but the same level of improvement is still to be made for the micro-wave reference unit. Before this work, the latter was directly adapted from the developments carried out for atomic fountain micro-wave clocks~\cite{Rosenbusch}. The modules currently used for atom interferometers are bulky and feature a high power consumption. They thus remain one of the last barriers to full mobility.\\ 
Our synthesizer results from a trade-off between mobility and high sensitivity measurement. We took into account the specific frequency response of atom interferometers~\cite{Cheinet}, as well as their main limitation due to ground vibrations~\cite{Merlet2}. Requirements on the residual phase noise and the frequency stability have been derived in the case of a Mach-Zender-like $\pi/2-\pi-\pi/2$ interferometer with 100 ms of interrogation time, as a guideline for the system specifications.\\
The paper is organized as follows: first we describe the main characteristics of our method and the system architecture we implemented. We then evaluate the performances of our set-up compared to the frequency standards operating at LNE-SYRTE in order to validate our design. We finally assess what sensitivity to acceleration an atomic interferometer  can achieve using our synthesizer, and compare this to the state-of-the-art.

\section{\label{sec:level1}DESIGN AND INTEGRATION}

\subsection{\label{sec:level2}Requirements for the operation of a Raman transition based atom interferometer}

Atom interferometers relying on stimulated Raman transitions employ a micro-wave signal to set the frequency difference between the two lasers at the hyperfine transition frequency of the atoms. In order to further simplify the design, we apply a method previously demonstrated in one of our gravimeters~\cite{Louchet}. It consists in using the same micro-wave reference signal during the cooling and detection stage (between the cooling and the repumping beams), and during the interferometer also (between the two Raman beams). The micro-wave synthesizer thus needs to feature a significant frequency agility. In the case of $^{87}$Rb atoms, its output signal frequency has to be changed from 6.567 GHz to 6.834 GHz in a few ms, between the cooling and the interrogation stage. In practice, the micro-wave frequency can be controled by two different techniques: either with an opto-electronic phase-locked loop of the two laser beams in the micro-wave domain~\cite{Santarelli, Kasevich}, or with side-band generation using light modulation~\cite{Bouyer, Lienhart}. In both cases, the micro-wave reference sets the phase difference between the two Raman lasers, which is crucial for the interferometer.\\ 
Our device is intended to feed an electro-optic modulator (EOM), so that the carrier and the generated side-band form the two laser frequencies that will address the atoms during the entire measurement sequence. The amount of micro-wave power sent to the EOM controls the amount of optical power transferred to the sideband. The proper operation of our interferometer actually requires for our synthesizer a maximum output power level in the order of 25~dBm~\cite{EOM}. In addition, most of atom interferometers use a micro-wave signal to select the internal quantum state of the atoms. In our case, a square micro-wave pulse of typically 5 ms at a fixed frequency of 6.834~GHz is used to select the $^{87}$Rb atoms in the state $|F=2; m_f=0\rangle $.\\
As a result our synthesizer features two channels. One used to perform the atomic state selection, will be denoted as ``selection channel" in the following. The second channel, controlling both the frequency and the phase between the two laser lines, will be denoted as "Raman channel". 

\subsection{\label{sec:level2}System architecture}

\begin{figure}[h]
       \includegraphics[width=7cm]{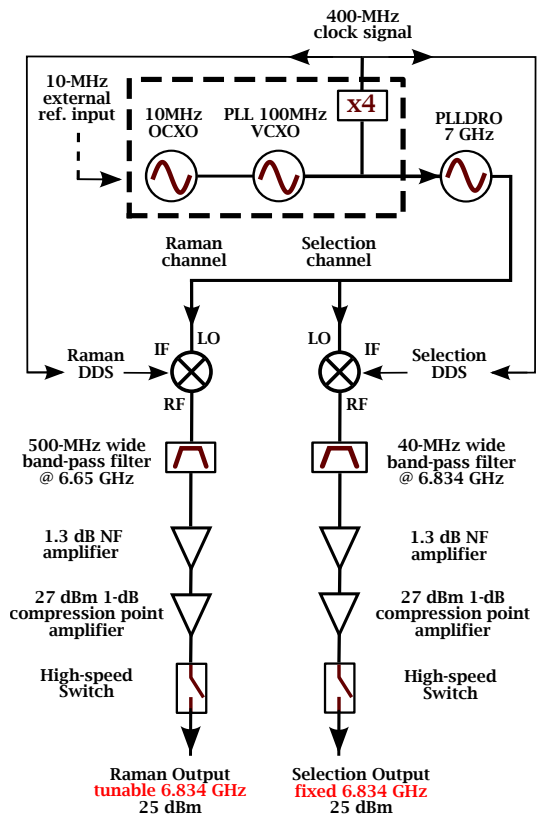}
    \caption{Schematic of our synthesizer. Two channels are derived from a single 7 GHz PLLDRO phase locked on a 100 MHz reference. The outputs supply 25 dBm for a fixed 6.834 GHz signal and for a tunable micro-wave signal to drive Raman transitions}
    \label{ArchiChaine}
\end{figure}

Both the Raman and the selection channels are created from a single source as shown on Fig.~\ref{ArchiChaine}. Our device relies on an initial custom-designed radio-frequency (RF) stage~\cite{ARE}. This subsystem is thermo-regulated and includes an oven controlled crystal oscillator (OCXO) reference working at 10~MHz. It is followed by a 100-MHz voltage controlled crystal oscillator (VCXO), internally phase locked on the OCXO via a 30-Hz bandwidth loop. Even if our device is intended to be used in free-running mode, this first stage can be phase-locked to a 10-MHz external reference when available. We used this option for the characterization of the phase noise level of the Raman channel. The RF module presents two outputs: one delivers a sinusoidal signal featuring +7~dBm at 400~MHz after internal multiplication, devoted to the synchronization of external devices. The second output provides a sinusoidal signal featuring +10~dBm at 100~MHz. The latter feeds a phase-locked loop dielectric resonant oscillator (PLLDRO) working at a fixed frequency of 7~GHz,  recently available commercially~\cite{Nexyn}. This component replaces the usual multiplication stage using non-linear transmission lines or step recovery diodes~\cite{Chambon}. The two channel signals are generated in an identical fashion from the same DRO output signal, by using double balance mixers and direct digital synthesis (DDS)~\cite{DDS} devices. The DRO output signal drives the local oscillator (LO) port of the mixers and the DDS device signal drives their intermediate frequency (IF) port. Each DDS synthesizer is clocked by the 400-MHz output signal of the RF initial stage, and controls the amplitude, the frequency and the phase of output signals. 
After the mixers, each channel is bandpass filtered to suppress the carrier and the spurious harmonics produced by the mixing process. Since the selection channel frequency is fixed, we use a 40-MHz bandpass filter centered on 6.834~GHz and specified to reject 60~dB at 7.0~GHz. As a significant frequency agility is required on the Raman channel, we use a 500-MHz bandpass filter centered on 6.65 GHz, specified to attenuate 50~dB at 7.0~GHz. The amplification on each of the two channels is performed using a low-noise amplifier (25~dB gain, 12~dBm, 1~dB compression point, 1.3~dB noise figure), followed by a high power amplifier (24~dB gain, 27.5~dBm, 1~dB compression point). Finally, two high-speed TTL-driven switches (100-ns switching speed, 70-dB attenuation by absorption) are used to control the two outputs of our device independently. \\
All the selected components are based on mature and commercially available technologies so as to benefit from compactness, lower power consumption, and minimal maintenance. These parameters have been confronted by the performances demanded by the state-of-the-art sensitivities of atomic measurements of acceleration. The complete micro-wave synthesizer described above fits in a two-litre physics package, and features a power consumption of about 30~W. The power consumption of the 10-MHz - 100-MHz - 7-GHz oscillators is 7.5~W. 

\section{\label{sec:level1}EXPERIMENTAL RESULTS}

\subsection{\label{sec:level2}Spurious harmonics}

In our atomic interferometer, micro-wave spurious harmonics create optical spurious sidebands. As these are not resonant with the atoms, this only leads to a decrease in the optical power. We have characterized at the output of the Raman channel the spurious harmonics generated by the mixing process, after their propagation through the filter and the amplification stages. We found the maximum ratio between the power levels of the carrier and the spurious harmonics, for a Raman DDS output signal power level of -6~dBm. The strength carried by the wave at the LO port is 6.5~dBm. In this case, we report in Tab.~\ref{tab:spurious} on the corresponding spurious harmonics at the output of the Raman channel for a frequency of the DDS device set at 166~MHz. Tab.~\ref{tab:spurious} shows that less than 1 part out of 10000 of the micro-wave power is lost, which is negligible. Furthermore, we characterized the isolation between the Raman and the selection channels: the power level of the crosstalk harmonics remains below -70~dBm, which is suitable for our application.

\begin{table}[h]
\caption{Spurious harmonics observed at the output of the Raman channel for a radio-frequency signal of 166 MHz and -6 dBm feeding the IF port of the mixer. In this case the side-band of interest at 6.834 GHz carries a power of 24.5 dBm.}
\label{tab:spurious}
\begin{ruledtabular}
\begin{tabular}{lllllll}
Harmonics & -4 & -3 & -2 & -1 & 0 & +1\\
\hline
Frequency (GHz) & 6.336 & 6.502 & 6.668 & 6.834 & 7.0 & 7.166\\
Power ratio (dBc) & $<$-100 & -43 & -41 & 0 & -49 & -51\\
\end{tabular}
\end{ruledtabular}
\end{table}

\subsection{\label{sec:level2}Accuracy and long-term drifts}

The accuracy of the measurement performed by atom interferometers based on stimulated Raman transitions, depends directly on the micro-wave reference signal's frequency accuracy~\cite{Louchet}. As an example, precise measurements performed by atomic gravimeters require a relative uncertainty of $10^{-9}$ for the micro-wave reference signal frequency. To determine the quality of this signal, we first evaluated the frequency accuracy of the 100-MHz output signal of the free-running RF stage. Then, we estimated the frequency stability of the PLLDRO output signal, locked on this free-running oscillator. We mix the signals of interest with the output signal of an auxiliary synthesizer, locked on an external frequency reference. This separate reference is driven by a hydrogen maser, and linked to a primary frequency standard running at the LNE-SYRTE laboratory. The relative frequency offset of the output frequency of our free-running synthesizer is in the order of $2.5 \times 10^{-10}$. We study its stability by monitoring the beatnote frequency at 7~GHz and we tested the warm-up of our device at room temperature. A duration of 30~min (resp. one hour and a half) is needed to reduce the output frequency relative offset below $1 \times 10^{-9}$ (resp. $2.5 \times 10^{-10}$), compared to its final value. In addition, the relative frequency stability of the RF-stage 100-MHz output signal due to temperature has been characterized by the manufacturer. It is below $1 \times 10^{-9}$ over the range [0 , $50\,^{\circ}{\rm C}$]. This allows a use of our device almost in this entire temperature range. 

\subsection{\label{sec:level2}Residual phase noise of the output signals}

\begin{figure}[h]
       \includegraphics[width=9cm]{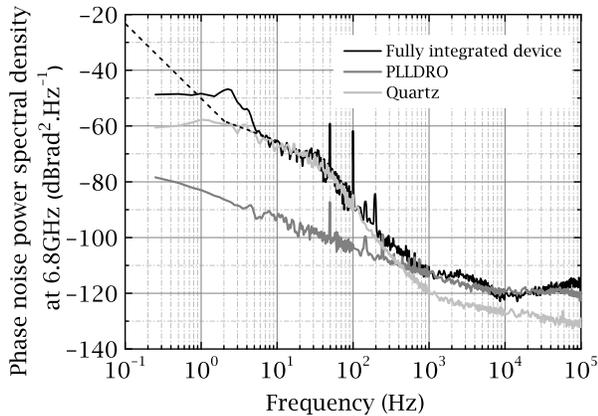}
    \caption{Power spectral density of the residual phase noise of the Raman channel output signal measured at 6.8~GHz (in black), of the quartz oscillator measured at 100~MHz and transposed at 6.8~GHz (in light grey), and of the PLLDRO (in grey). Below 2~Hz the doted line represents the measured behaviour of the free-running quartz reference transposed at 6.8~GHz. }
    \label{phase}
\end{figure}

We measure the phase noise power spectral density (PSD) of the Raman channel output signal of our micro-wave reference, to assess what would be the sensitivity of the atomic measurement. We thus compare the Raman channel output signal of our fully integrated set-up, with respect to a second reference synthesizer output signal at 6.8~GHz. The phase noise level of the latter was measured separately to be lower than our specifications~\cite{LeGouet}. For the present measurement, our system is phase-locked on this reference. The two signals are fed in quadrature into a phase detector, and the noise of the resulting voltage is measured by a FFT. Prior to the measurement of the phase noise PSD, we performed a calibration step to relate voltage noise to phase noise, with a precision assessed to be better than 5~$\%$. The corresponding phase noise PSD is shown in black on Fig.~\ref{phase}. The grey (resp. light grey) curve represents the PSD of the residual phase noise of the PLLDRO measured at 7~GHz (resp. of the quartz measured at 100~MHz and transposed at 6.8~GHz). 

For these two last measurements, we compared two identical devices, both phase locked on a common reference signal following the method described above. It shows that for Fourier frequencies below 400~Hz, the contribution of the quartz oscillator to the total residual phase noise dominates. Above this frequency, the phase noise performance of our synthesizer is limited by the contribution of the PLLDRO. On the black curve, the increase in the phase noise level for Fourier frequencies above 30~kHz is imputed to the synthesizer that we used as a reference. The bump observed for frequencies between 1~kHz and 10~kHz is attributed to the Raman DDS unit. Aside from this, it is worth pointing that the bandwidth of the PLLDRO's servo loop on the RF stage features no noticeable contribution to this curve. The bump around 30~Hz is due to the bandwidth of the phase-lock loop of the 100-MHz VCXO on the 10-MHz OCXO of the RF stage. The shape of the black curve below 6 Hz is due to the bandwidth of the phase-lock loop of our primary quartz oscillator on the reference. A specific study of the phase noise PSD for Fourier frequencies below 10~Hz has been carried out. We mix the 100-MHz output signal of our system with the 100-MHz output signal of the previous reference. In this case, the reference synthesizer is phase-locked on our apparatus via a 0.05~Hz bandwidth loop. The residual phase noise transposed at 6.8~GHz is displayed on Fig.~\ref{phase} by the dotted straight lines. Below 2~Hz, the slope of the spectrum is determined to be -2.7.

\subsection{\label{sec:level2}Calculated impact on the sensitivity of the atomic measurement}

The sensitivity of an atomic interferometer to different sources of noise can be characterized by the Allan variance~\cite{allan} of the interferometric phase fluctuations. For long enough averaging time it has been shown~\cite{Cheinet} that this Allan variance is given by:

\begin{equation}
\label{SigmaPhi}
  \sigma_{\phi}^2(\tau) = \frac{1}{\tau} \sum_{n=1}^{\infty}\vert H(2 \pi n f_c) \vert^2 S_\phi(2 \pi n f_c)
\end{equation}

\begin{figure}
       \includegraphics[width=9cm]{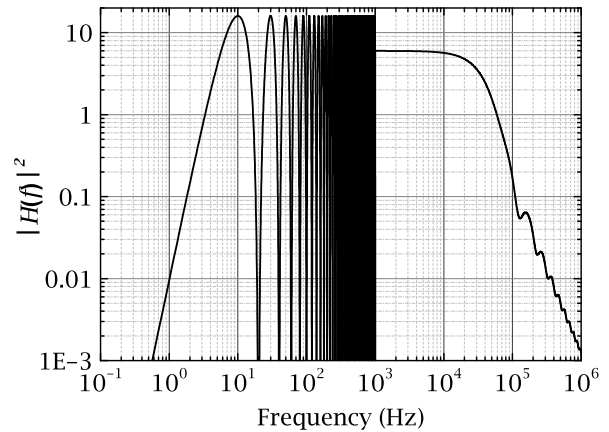}
    \caption{Sensitivity function to Raman laser phase noise for an atomic accelerometer based on a $\pi/2-\pi-\pi/2$ sequence and for $\pi/2$ pulse duration $\tau_R$ = 10 $\mu$s, interrogation time 2T~=~100~ms. Below 1 kHz it shows an oscillation with period of T+2$\tau_R$. For Fourier frequencies above 1~kHz only the mean value is represented}
    \label{TransferFunc}
\end{figure}

where $\tau$ is the averaging time, H is the sensitivity function of the interferometer~\cite{Cheinet} (represented in Fig.~\ref{TransferFunc}), $f_c = \frac{1}{T_c}$ the cycling frequency, and $S_\phi$ is the PSD of the phase difference between the two Raman lasers. Eq.~\ref{SigmaPhi} shows that the PSD of the residual phase noise of our micro-wave signal $S_\phi$, and the sensitivity function of the atomic interferometer, enable us to estimate the contribution of our set-up to the sensitivity of the atomic measurement of acceleration. The shape of the sensitivity function, displayed on Fig.~\ref{TransferFunc}, is determined by the time-domain structure of a measurement cycle~\cite{Cheinet}. It features three main time scales: the $\pi/2$ Raman pulse duration $\tau_R$, the interrogation time 2T and the cycling time of the measurement T$_{c}$. The finite length of the Raman light pulses acts as a natural first order low-pass filter, whose cut-off frequency is $1/(2\tau_R)$. In contrast with atomic clocks, the three-pulse interrogation sequence performs a high-pass filter, whose cut-off frequency is $1/(2T)$. This drives the sensitivity to residual phase noise down to zero as the frequency decreases. Furthermore, the operation of atomic inertial sensors is characterized by dead-times between two consecutive interferometric measurements. They are necessary to prepare the atomic source at the input of the interferometer and to read-out its output. This pulsed mode of operation induces an aliasing effect, that ultimately limits the sensitivity of our instruments due to the noise sampled at the multiple harmonics of the cycling frequency (Eq.~\ref{SigmaPhi}). This is a direct analog of the Dick effect encountered by atomic clocks~\cite{dick}. \\
The contribution of the atomic phase fluctuations to acceleration sensitivity is given by~\cite{Louchet}:

\begin{equation}
\label{SigmaAcc}
  \sigma_a(\tau)  = \frac{\sigma_{\phi}(\tau)}{k_\text{eff} T^2} = \frac{1}{k_\text{eff} T^2 \sqrt{\tau}} \sqrt{\sum_{n=1}^{\infty}\vert H(2 \pi n f_c) \vert^2 S_\phi(2 \pi n f_c)}
\end{equation}

where $k_\text{eff}$ is the effective wave-vector. This shows that, in the case where the sensitivity to acceleration is limited by the residual phase noise of our synthesizer, the Allan variance should behave as $1/\sqrt{\tau}$ for long enough integration time. We extrapolate from this behaviour a short-term atomic phase stability per shot $\sigma_{\phi}(T_c)$.

\begin{figure}[h]
       \includegraphics[width=9cm]{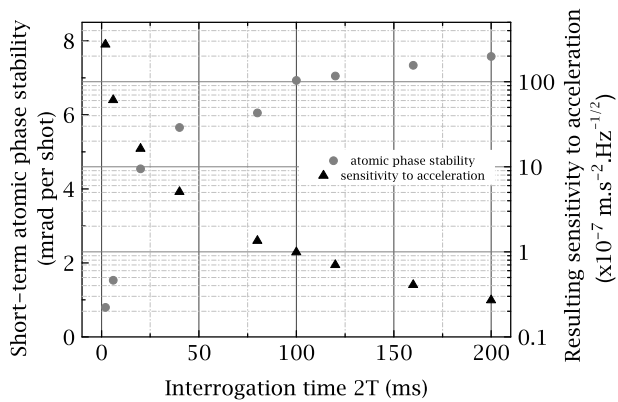}
    \caption{Evolution of the corresponding short-term sensitivity with the total interrogation time 2T. We express it in atomic phase fluctuation per shot, and in acceleration sensitivity for a cycling time kept constant at 333~ms.}
    \label{experiset}
\end{figure}

In the following, we report on the calculation of the impact of our compact synthesizer on the short-term performance of the atomic interferometer, for a wide range of values of $\tau_R$, 2T and T$_c$. 
Fig.~\ref{experiset} shows the results of the calculations of Eq.~\ref{SigmaPhi} and Eq.~\ref{SigmaAcc} for increasing interrogation time, keeping the cycling time constant at 333~ms and $\tau_R$ equal to 10 $\mu$s. The values presented in this figure are the short-term statistical uncertainties on the atomic phase, due to random fluctuations of the phase of our micro-wave reference signal. Fig.~\ref{experiset} is, under the assumptions of Eq.~\ref{SigmaPhi}, a statistical representative description of the effect of our synthesizer, estimated from a single realization of the micro-wave phase noise. The short term stability increases rapidly until 2T reaches 40~ms, but then does not extend much further for longer interferometer durations. It remains below 8 mrad per shot even at 200~ms total interrogation time. In the mean time, the sensitivity to acceleration increases quadraticaly, as seen in Eq.~\ref{SigmaAcc}. This stays true as long as the slope of the PSD of the residual phase noise of our set-up remains greater than -4 for Fourier frequencies below 10~Hz. In particular, we evaluate a resulting atomic phase stability of 7~mrad at one shot for typical interferometric parameters ($\tau_R$ = 10 $\mu$s, 2T~=~100~ms and T$_c$ = 333~ms). This is equivalent to having a signal to noise ratio of 142. In this case, we calculated that our device would allow for a sensitivity of the measurement at the $10^{-7}$m.s$^{-2}$.Hz$^{-1/2}$ level. This result is in the order of the best sensitivities reported in the laboratory with such experimental parameters~\cite{LeGouet}. 
The variation of $\tau_R$ between 1 $\mu$s and 1~ms while 2T and T$_c$ are kept constant does not significantly change the result. This means that the contribution of the residual phase noise of our synthesizer, carried by Fourier frequencies above 1~kHz (chiefly due to the PLLDRO), is negligible. The study of the impact of the PLLDRO output signal phase noise  gives a short term stability of 0.7~mrad per shot, corresponding to a sensitivity to acceleration of $1 \times 10^{-8}$ m.s$^{-2}$.Hz$^{-1/2}$ (for $\tau_R$ = 10 $\mu$s, 2T~=~100~ms and T$_c$ = 333~ms). This implies that the PLLDRO is not limiting the short-term atom interferometric inertial measurements, for any temporal parameters. 
In addition, we calculated that the greatest spectral contribution of the PSD to residual atomic phase fluctuations is carried by Fourier frequencies in the range of 1~Hz to 30~Hz. In our case, the 10-MHz OCXO quartz oscillator is the limiting element in this area. As a result, the influence of our micro-wave reference on the sensitivity of the interferometric measurement can be further decreased by the use of a better quartz oscillator. 

The total interrogation time is varied significantly regarding the type of application of the atomic interferometer. However, the practical operation of such sensors is always characterized by unavoidable dead-times between two consecutive measurements. Another way of presenting our results is thus to show the effect of the residual phase noise due to our microwave reference signal on the sensitivity to acceleration for different values of 2T, while keeping the dead-time equal to 250~ms (Tab.~\ref{tab:table2}). 

\begin{table}[h]
\caption{\label{tab:table2}Corresponding short-term contribution to atomic phase fluctuations that our micro-wave reference signal would induce, for various operating regimes of interferometer. The dead time is fixed at 250~ms}
\begin{ruledtabular}
\begin{tabular}{cccc}
T$_c$ (ms) & 2T (ms) & $\sigma_{\Phi}$ (mrad/shot) & $\sigma_a$ (10$^{-7}$m.s$^{-2}$.Hz$^{-1/2}$)\\
\hline
256 & 6 & 1.6 & 54\\
300 & 50 & 6.3 & 3.4\\
350 & 100 & 7.0 & 1.00\\
570 & 320 & 8.6 & 0.2\\
850 & 600 & 11.1 & 0.07\\
1250 & 1000 & 15.1 & 0.04\\
\end{tabular}
\end{ruledtabular}
\end{table}

We show that even if our apparatus was designed for a 100-ms interferometer, it will not degrade the inertial measurement sensitivity for longer interferometric times. In particular, the highest sensitivity to acceleration at the level of $4.2 \times 10^{-8}$ m.s$^{-2}$.Hz$^{-1/2}$ has been reported in Ref~\cite{Hu}. Using the experimental parameters from the above reference ($\tau_R$ = 14 $\mu$s, 2T = 600 ms and T$_c$ = 1 s), we calculate that our synthesizer would allow us to reach a sensitivity to acceleration as low as $0.8 \times 10^{-8}$ m.s$^{-2}$.Hz$^{-1/2}$.

\section{\label{sec:level1}CONCLUSION}

We have demonstrated the realization of a two-litre compact, 30-W micro-wave frequency reference. Our device is intended to support the field operation of atom interferometers based on stimulated Raman transitions. It benefits from the use of an integrated commercial PLLDRO, and its power consumption could be further decreased to 20~W by removing the high-power amplifiers required by our specific experimental set-up.\\ 
Designed for atom gravimeters of 100-ms long interrogation time, our micro-wave synthesizer has been characterized to allow for sensitivity to acceleration at the level of $10^{-7}$m.s$^{-2}$.Hz$^{-1/2}$. This result is in the order of the best laboratory-recorded performances. Furthermore, we have calculated that the residual phase noise level of our micro-wave reference would even enable us to reach state-of-the-art sensitivities to acceleration for longer interrogation time~\cite{Hu}.\\ 
This work opens a doorway for the transfer of light pulse atom interferometry from laboratory towards field and on-board operation, for mobile gravimetry surveys or inertial navigation.

\begin{acknowledgments}
We would like to thank Franck Pereira Dos Santos for useful discussions, Laurent Volodimer and Jos\'{e} Pinto for the supply of the DDS device, Peter Rosenbusch and Indranil Dutta for careful reading of the manuscript. We also would like to thank the Institut Francilien pour la Recherche sur les Atomes Froids (IFRAF), the Agence Nationale pour la Recherche (ANR contract ANR - 09 - BLAN - 0026 - 01) in the frame of the MiniAtom collaboration for funding this work. J.L. would like to thank UPMC and College des Ingenieurs, in the frame of the program "Science and Management" for supporting his work.
\end{acknowledgments}


\begin{thebibliography}{10}
\bibitem{DDS} Selection DDS is AD9959 and Raman DDS is AD9852. The latter is used in sweep mode and is overclocked. Gravity measurements require 48-bit resolution and phase continuity for the frequency chirp that compensates for the Doppler shift~\cite{Louchet}.
\end{thebibliography}
\end{document}